\documentclass[aps,twocolumn,showpacs,preprintnumbers,floatfix]{revtex4}
\usepackage[T1]{fontenc}
\usepackage{times}
\usepackage{mathptmx}
\usepackage{bm}
\usepackage{graphicx}
\usepackage{dcolumn}
\usepackage{tabularx}

\begin{document}

\preprint{APS/123-QED}
\title{Magnetotransport Study of the Canted Antiferromagnetic Phase in
Bilayer $\nu=2$ Quantum Hall State}

\author{A. Fukuda}
\email{fukuda@scphys.kyoto-u.ac.jp}
\author{A. Sawada}
\affiliation{Research Center for Low Temperature and Materials Sciences, Kyoto 
University, Kyoto 606-8502, Japan}
\author{S. Kozumi}
\author{D. Terasawa}
\author{Y. Shimoda}
\author{Z. F. Ezawa}
\affiliation{Graduate School of Science, Department of physics, Tohoku University, Sendai  980-8578, Japan}
\author{N. Kumada}
\author{Y. Hirayama}
\affiliation{NTT Basic Research Laboratories, NTT Corporations, 3-1 Morinosato-Wakamiya, Atsugi 243-0198, Japan}
\date{\today}

\begin{abstract}
Magnetotransport properties are investigated in the bilayer quantum Hall state at the total filling factor $\nu=2$. We measured the activation energy elaborately
as a function of the total electron density and the density difference between the two layers.  Our experimental data demonstrate clearly the emergence of the canted antiferromagnetic (CAF) phase between
the ferromagnetic phase and the spin-singlet phase.  
The stability of the CAF phase is discussed by the comparison between
experimental results and theoretical calculations using a Hartree-Fock
approximation and an exact diagonalization study. 
The data reveal also an intrinsic structure of  the CAF phase divided into two regions according to the dominancy between the intralayer and interlayer correlations.

\end{abstract}

\maketitle




\section{INTRODUCTION}

Two-dimensional electron gas provides us with a simplest low-dimensional condensed matter system having the spin degree of freedom. It has a conceptual link to 
the unconventional superconductivities\cite{Maeno}, Bose-Einstein
condensation in two dimensions\cite{Gorlitz,Schweikhard} and superfluidity of $^3$He film\cite{Davis}. 
However, more fascinating phenomena are anticipated by introducing an additional degree of freedom to the two-dimensional electron systems.

In the last decade, many efforts were devoted to study bilayer quantum Hall (QH) systems\cite{DasSarma_book, Ezawa_book}, which have an additional layer degree of freedom
associated with the third dimension. The layer degree of freedom is regarded
as a virtual one-half spin system and named 'pseudospin'. Although the total filling factor $\nu =1$ QH system has only one phase without in-plane magnetic field, the $\nu =2$ QH system has a variety of quantum phases related
to a combination of the spin and the pseudospin. 
At $\nu =2$, in a naive one-body picture, a phase transition occurs because of the  competition between the tunneling energy $%
\Delta _{\mathrm{SAS}}$ and the Zeeman energy $\Delta _{\mathrm{Z}}$. As a
result, two phases occur: one is the spin-ferromagnet and
psuedspin-siglet phase (F phase) for $\Delta _{\mathrm{SAS%
}}<\Delta _{\mathrm{Z}}$ and the other is the spin-singlet and
pseudospin-ferromagnet phase (S phase) for $\Delta _{%
\mathrm{SAS}}>\Delta _{\mathrm{Z}}$. 
The F phase consists of two single-layer $\nu =1$ QH systems. 
The S phase has an interlayer phase coherence due to the
interlayer Coulomb interaction. Between these two phases, a novel canted
antiferromagnet phase (CAF phase) has been 
argued to emerge\cite{Zheng,Das}. 

The first experimental indication of a new phase in the $\nu=2$ bilayer QH system
was given by 
inelastic light scattering spectroscopy by Pelligrini \textit{et al}\cite{Pelligrini_PRL}.
They also observed mode softening signals indicating  second-order phase transitions\cite{Pelligrini_Science}.
The new phase was theoretically identified as the CAF phase by Das Sarma \textit{et al}\cite{Zheng,Das}. 
They obtained the phase diagram in the $\Delta_{\mathrm{SAS}}$-$d$ plane 
based on a time-dependent Hartree-Fock (HF) analysis,
where $d$ is the layer separation. An effective spin theory\cite
{Demler_Bose,Yang}, a Hartree-Fock-Bogoliubov approximation\cite
{Shimoda} and an exact diagonalization (ED) study
\cite{Schliemann} followed to improve the phase diagram by more precise
calculations. Effects of the density imbalance on the CAF phase were also discussed and several different phases are expected to emerge\cite{MacDonald_Rajaraman,Brey}. Recently the
phase diagram in the $\sigma-n_{\mathrm{T}}$ plane was constructed\cite%
{Ezawa_can}, where $n_{\mathrm{T}}$ is the total electron density and $%
\sigma $ is the density imbalance between the two layers. 
Though capacitance spectroscopy\cite{Khrapai} as well as magnetotransport
measurements\cite{Sawada_PRL,Sawada_PRB,Kumada,Geer} were carried out,
experimental studies on the CAF phase remain less systematic
than theoretical works.  

This paper is organized as follows. 
In Section II our experimental setup is described. 
In Section III we report the results of elaborate megnetotransport experiments performed to quest for essential features of the CAF phase. Activation energy measurements as a function of $n_{\mathrm{T}}$ and $\sigma $ show clearly that there exist three phases to be identified as the F, S and CAF phases. 
In Section IV we present the phase diagram in the $\sigma $ - $n_{\mathrm{T}}$ 
plane. We have demonstrated that the layer density imbalance causes new quantum states, \textit{i.e.} two regions in the CAF phase, which have never been predicted theoretically.  We also try to figure out the intrinsic spin and pseudospin structures of new regions.
In Section V the phase diagram of the $\nu =2$ bilayer QH system in the $n_{\mathrm{T}}$-$\Delta _{\mathrm{SAS}}$ plane is presented. Theoretical considerations by the HF approximation and the ED study are made. The ED analysis gives a quantitative support to the emergence of the CAF phase. 
In Section VI, we comment on the comparison between our results and several theoretical works.


\section{EXPERIMENTS}


We used a sample consisting
of two GaAs quantum wells of $20$\thinspace nm in width separated by a $3.1$%
\thinspace nm-thick Al$_{0.33}$Ga$_{0.67}$As barrier. The sample was grown
by molecular-beam epitaxy. Si-modulation doping is carried out only on the
front side of the double-quantum-well (DQW) structure and electrons in the back side of the DQW is fully
field-induced by applying a positive bias to the underlying $n^{+}$-GaAs
gate \cite{Muraki_sample}. The tunneling energy $\Delta _{\mathrm{SAS}}$ is
11\thinspace K, the layer separation $d$ is 23.1\thinspace nm, and the low
temperature mobility at $n_{\mathrm{T}}=1.0\times 10^{11}\mathrm{cm}^{-2}$
is $1.0\times 10^{6}\mathrm{cm}^{2}/\mathrm{Vs}$. We can control $n_{\mathrm{%
T}}$ up to $3.0\times 10^{11}\mathrm{cm%
}^{-2}$ and the density imbalance parameter $\sigma $ from 0 at the balanced
configuration to $\pm 1$ at the monolayer configuration continuously by
applying the front- and back-gate voltages. The density imbalance parameter is
defined by $\sigma \equiv(n_{\mathrm{f}}-n_{\mathrm{b}})/(n_{\mathrm{f}}+n_{%
\mathrm{b}})$, where $n_{\mathrm{f}}$ ($n_{\mathrm{b}}$) is the electron
density in the front (back) layer. 
To measure resistances, standard low-frequency ac lock-in techniques
were used with a current of 20\thinspace nA and a frequency of 16.6 Hz.
Throughout measurements, magnetic field is applied perpendicular to the two-dimensional plane.

\section{ACTIVATION ENERGY MEASUREMENTS}

\begin{figure}[t]
\begin{center}\leavevmode
\includegraphics[width=\linewidth]{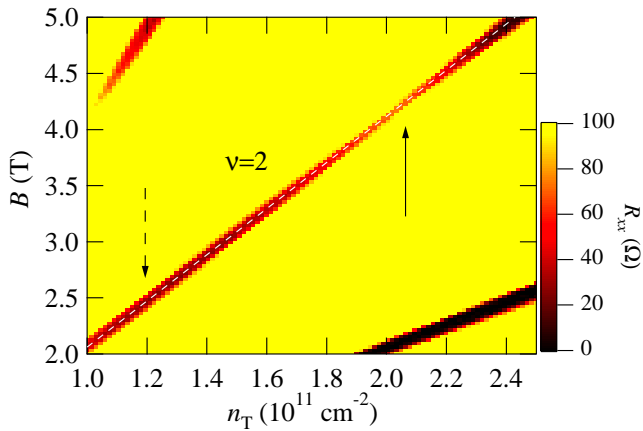}
\caption{Image plot of the magnetoresistance $R_{xx}$ around the $\nu=2$ QH state as a function of the total density $n_{\rm T}$ and the magnetic field $B$ at a temperature of 1.11 K. The white dashed line is located just on the $\nu=2$ filling. }
\label{DM0}
\end{center}
\end{figure}

 First, we investigate the bilayer $\nu=2$ QH states in the balanced density 
configuration ($\sigma=0$). Figure \thinspace\ref{DM0} shows $R_{xx}$ as a function of the 
 magnetic field $B$ and $n_{\mathrm{T}}$, where $n_{\mathrm{T}}$ is scanned while keeping $n_{\mathrm{f}}=n_{\mathrm{b}}$.
Dark regions represent smaller $R_{xx}$ and thus QH states.
The central black region around the white dashed line running from lower left to upper right in the figure indicates the $\nu=2$ QH state. The width of the black region is related to the stability of the QH state. With increasing $n_{\mathrm{T}}$, the width increases to the point indicated by the dashed arrow in Fig. \thinspace\ref{DM0}, then decreases to the one indicated by the solid arrow, and increases again. This implies that the $\nu=2$ QH state becomes stable, then less stable and stable again as a function of $n_{\mathrm{T}}$.

\begin{figure}[t]
\includegraphics[width=0.98\linewidth]{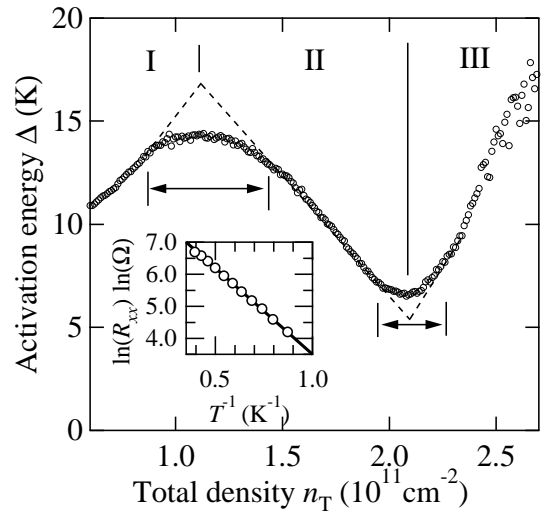}
\caption{Activation energy $\Delta$ as a function of the total density $n_{\mathrm{T}}$ in the balanced configuration. Three different regions are found and
labeled I, II and III. See details in text. Inset: Arrehnius plot of the
magnetoresistance $R_{xx}$ at $n_{\mathrm{T}} = 1.66\times10^{11} \mathrm{cm}%
^{-2}$. }
\label{delta_nt}
\end{figure}

 To clarify this fact more, we measured the $n_{\mathrm{T}}$ dependence of the activation energy $\Delta$, which  was determined from the slope of the Arrehnius plot of the longitudinal
resistance $R_{xx}$, $R_{xx}=R_{0}\mathrm{exp}\left( -\frac{\Delta }{
2T}\right) $, where $T$ is the temperature.  Figure \ref{delta_nt} shows the activation energy as a function of $n_{\mathrm{T}}$ in the balanced configuration ($\sigma =0$). In the
beginning, $\Delta $ gradually increases for the increment of $n_{\mathrm{T}%
} $ from $0.6\times 10^{11}\mathrm{cm}^{-2}$ to $1.1\times 10^{11}\mathrm{cm}%
^{-2}$. After crossing the maximum, $\Delta $ decreases to the minimum at $n_{\mathrm{T}}=2.1\times 10^{11}\mathrm{cm}^{-2}$. 
 Finally, $\Delta $ steeply increases after crossing the minimum point. 
This figure indicates that there are, at least, three phases in the balanced
configuration in the $\nu =2$ QH state. We named them the phases I, II and
III from the low $n_{\mathrm{T}}$ to high $n_{\mathrm{T}}$ region as in
Fig.\thinspace\ref{delta_nt}. According to a number of theories mentioned above, the phases I, II and III should correspond to the S, CAF and F phases, respectively. It should be noted that $\Delta $ changes smoothly from the region I to II, and from II to III. This indicates that both phase transitions are not first order, as agrees well with the theoretical results\cite{Zheng,Das,Ezawa_can}. 
It is also consistent with the light-scattering experimental result 
\cite{Pelligrini_Science}.
Although the exact phase transition point is not clear because of the smooth change, it would exist around  $n_{\mathrm{T}}$ that gives the local maximum or minimum, probably in the range shown as the two-headed arrows in Fig.\thinspace\ref{delta_nt}. Hereafter we adopt the point that gives the local maximum or minimum of $\Delta $ as a representative phase transition point.

\begin{figure}[t]
\begin{center}\leavevmode
\includegraphics[width=\linewidth]{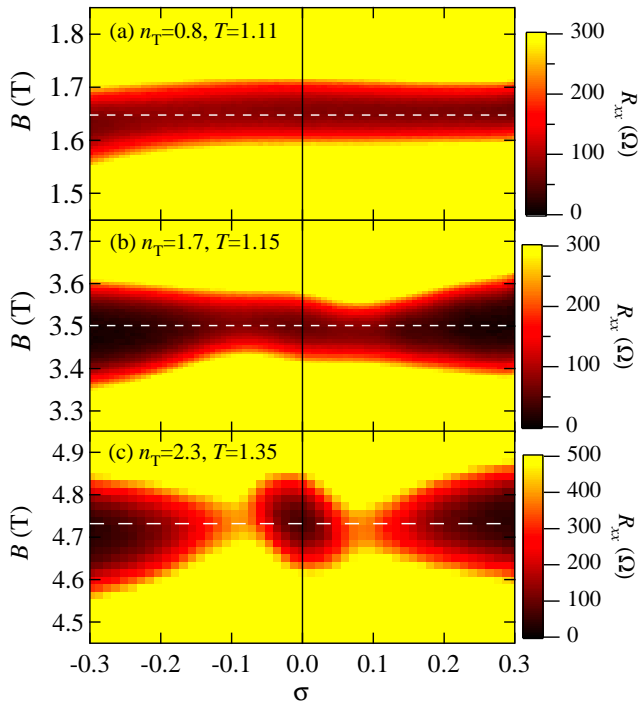}
\caption{Image plots of the magnetoresistance $R_{xx}$ around the $\nu=2$ QH state as a function of the density imbalance $\sigma$ and the magnetic field $B$ at the fixed total density $n_{\rm T}$. Each $n_{\rm T}$ ($10^{11} {\rm cm}^{-2}$) and temperature (K) are shown in the graph. The solid line indicates $\sigma=0$. The white dashed lines are located just on the $\nu=2$ filling.}
\label{SM0}
\end{center}
\end{figure}

To confirm the identification of these three phases at $\sigma =0$, 
we also investigated the stability of the $\nu=2$  QH states against the layer density imbalance $\sigma$ 
through the magnetotransport measurements in Fig.\thinspace\ref{SM0}. 
We present image plots of the magnetoresistance $R_{xx}$ by
changing the magnetic field $B$ and the layer density
imbalance $\sigma$. The black region represents the
well-developed $\nu =2$ QH states. We display three typical patterns of 
$R_{xx}$ for three values of  $n_{\rm T}$.  We remark the following characteristic features of the pattern as $|\sigma|$ is increased: 
(a) At $n_{\rm T}=0.8 \times 10^{11} \thinspace{\rm cm}^{-2}$ (region I at $\sigma=0$), the
width of the stable region is almost constant; 
(b) At $n_{\rm T}=1.7 \times 10^{11} \thinspace{\rm cm}^{-2}$ (region II at $\sigma=0$), it becomes
narrower and then wider slightly; 
(c) At $n_{\rm T}=2.3 \times 10^{11} \thinspace{\rm cm}^{-2}$ (region III at $\sigma=0$), it becomes narrower and
then wider drastically.

\begin{figure}[t]
\includegraphics[width=1\linewidth]{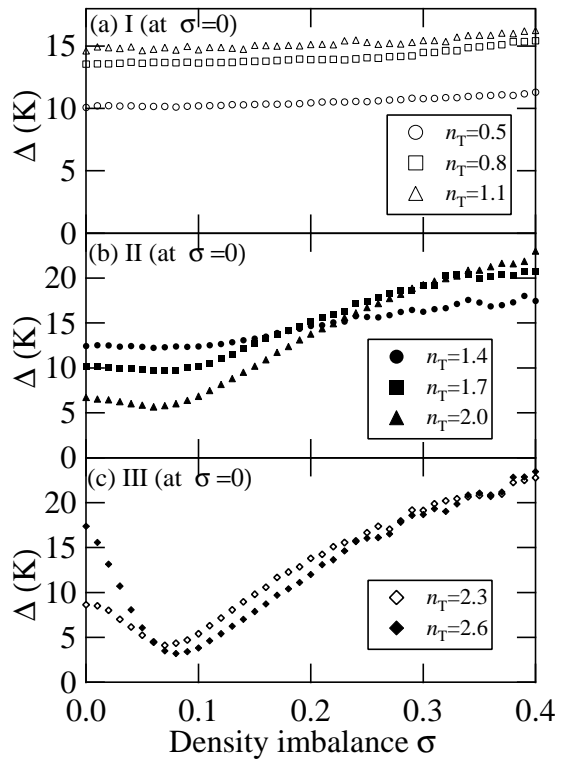}
\caption{Activation energy $\Delta $ as a function of the density imbalance $\protect\sigma $ at the fixed total density $n_{\mathrm{T}}$. 
Data in (a), (b) and (c) are for $n_{\mathrm{T}}$ in the phase I, II and III at $\sigma = 0$, respectively.
The unit of $n_{\mathrm{T}}$ noted in the graphs is $\times 10^{11}\mathrm{cm}^{-2}$. }
\label{delta_sigma}
\end{figure}

To investigate the stability of these three phases against the density imbalance further, in a similar way we did for $n_{\mathrm{T}}$, we also carried out the activation energy measurements as a function of $\sigma$. Results in each region are shown in Fig.\thinspace\ref{delta_sigma}. Although the identification of the phase at large density imbalance is important, about which we will discuss later,  here we focus on the properties in the vicinity of $\sigma =0$. 
In the phase I, the activation energy $\Delta $ is almost constant or gradually increases 
as $\sigma$ is increased. It indicates that the phase I at $\sigma =0$ is robust against the density imbalance. 
This fact supports that the phase I corresponds to the S phase because the S phase is stabilized by the interlayer correlation. On the other hand, in the phase III near $\sigma =0$, $\Delta $ steeply decreases for the initial increment of $\sigma $.  It indicates that the phase III at $\sigma =0$ is feeble against the density imbalance. This fact  points out that the phase III accords to the F phase because the F phase consists of two single-layer $\nu=1$ QH states\cite{Sawada_PRL,Sawada_PRB,Kumada}. In the phase II, $\Delta $ slightly decreases or is almost constant for small density imbalance. Taking account of the dependence of $\Delta $ on $n_{\mathrm{T}}$ and $\sigma $, the phase II is concluded to have properties quite different from the phases I and III.

\section{PHASE DIAGRAM IN THE $\sigma-n_{\mathrm{T}}$ PLANE}

\begin{figure}[t]
\includegraphics[width=0.95\linewidth]{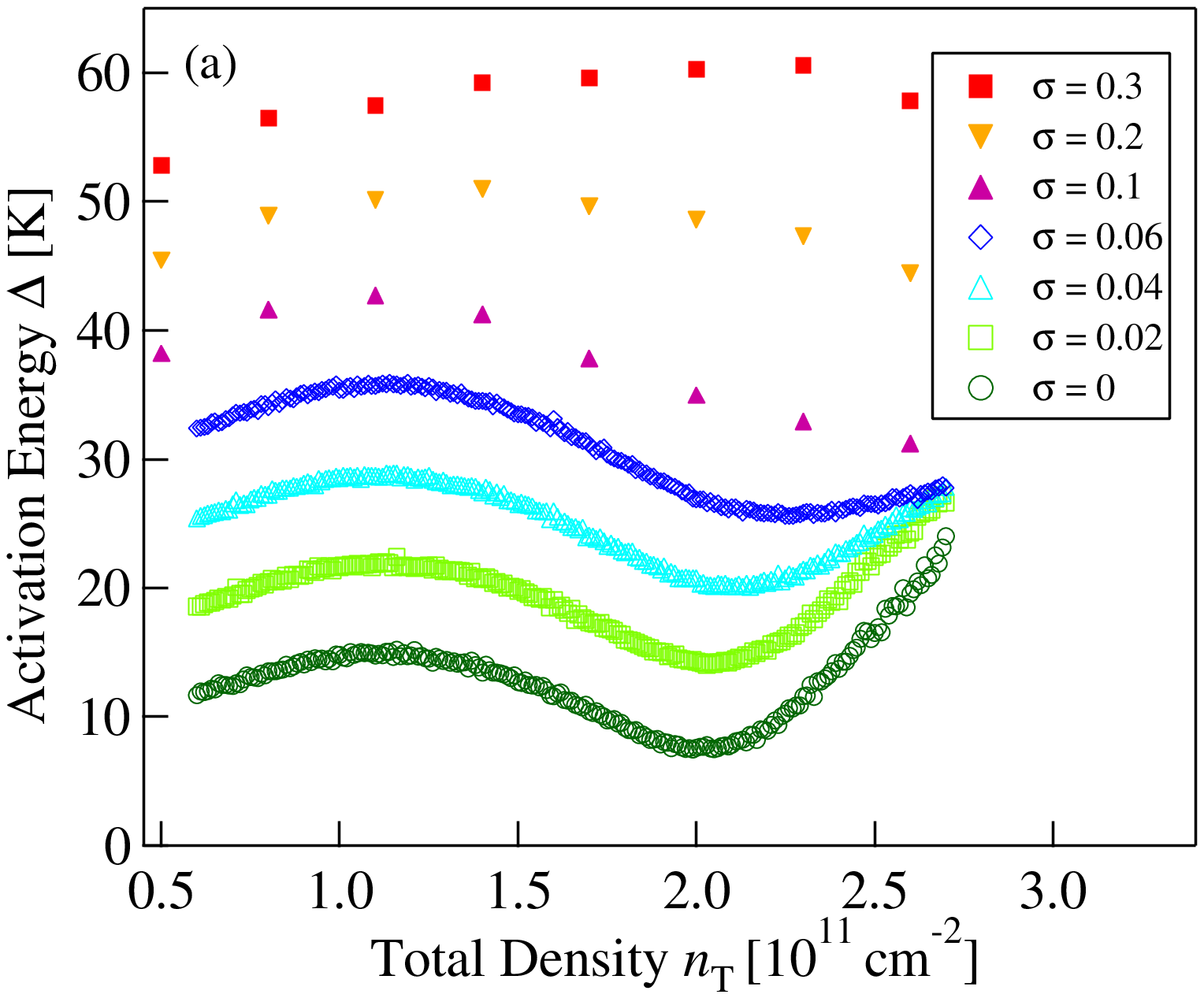}
\includegraphics[width=0.94\linewidth]{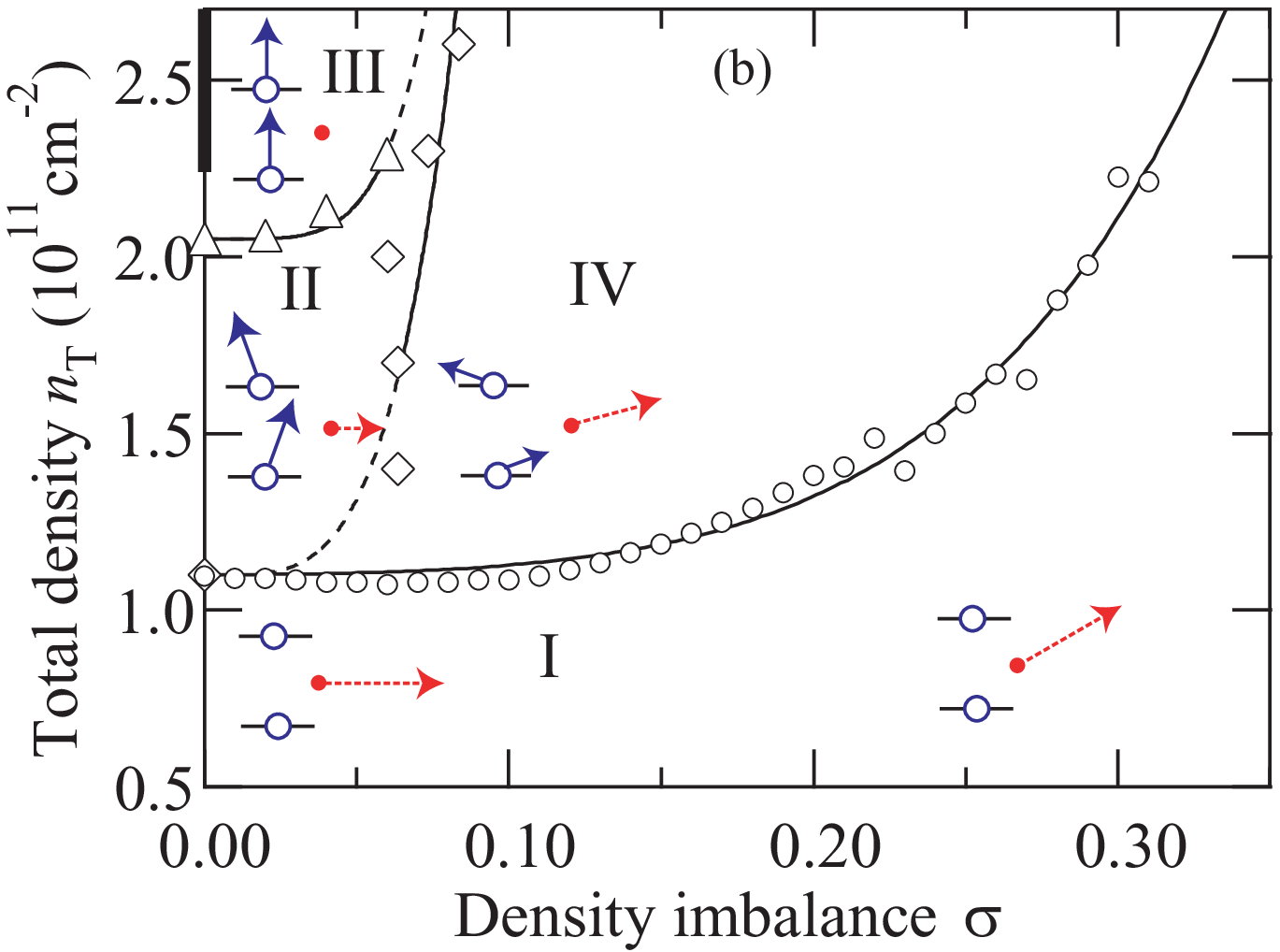}
\caption{
(a) Activation energy $\Delta $ as a function of the total density $n_{\mathrm{T}}$ for several density imbalance $\sigma$. The $\sigma$ for each data is listed in the graph.  The $\Delta$ at $\sigma = 0$ has a correct scale and each trace for $\sigma \neq 0$ is shifted by 7 K for easy to see. (b) 
Phase diagram in the $\protect\sigma $-$n_{\mathrm{T}}$ plane. Open triangles (circles)
are the $n_{\mathrm{T}}$'s that give the local minima (maxima) of 
the activation energy at fixed $\sigma$. 
The fat solid line just on the vertical axis for $n_{\mathrm{T}}>2.24\times 10^{11}\mathrm{cm}^{-2}$ is a
theoretically predicted F phase, which would be stable only at the balanced point but for an impurity effect.
Open diamonds give the II-IV region boundary.  
Solid lines are guides for eyes for region boundaries and dashed lines represent ambiguous boundaries.
The predicted structures of the spin (blue solid arrow) in the front and back layers and the total pseudospin (red dashed arrow) are also illustrated.}
\label{sigma_nt}
\end{figure}

We next construct the phase diagram  in the $\sigma $-$n_{\mathrm{T}}$ plane [Fig.\thinspace\ref{sigma_nt}(b)]. 
At $\sigma=0$ the CAF-F (S-CAF) phase boundary is given 
by the local minimum (maximum) of $\Delta$ in Fig.\thinspace\ref{delta_nt}.
We made similar measurements at various values of $\sigma $. 
Several data are displayed in Fig.\thinspace\ref{sigma_nt}(a).  
As $\sigma$ is increased, the $n_{\mathrm{T}}$ that gives the maximum of $\Delta$, the S-CAF phase boundary, stays almost constant when $\sigma<0.1$ and then moves towards the larger $n_{\mathrm{T}}$ side when $\sigma>0.1$.
On the other hand, the point that gives the minimum of $\Delta$, the CAF-F phase boundary, shifts to larger $n_{\mathrm{T}}$ for small increase in $\sigma$, and disappear when $\sigma>0.1$.

We plot the set of these values ($\sigma$,$n_{\mathrm{T}}$) as open triangles (circles) in Fig.\thinspace\ref{sigma_nt}(b), 
which gives the CAF-F (S-CAF) phase boundary.

\begin{figure}[t]
\includegraphics[width=1\linewidth]{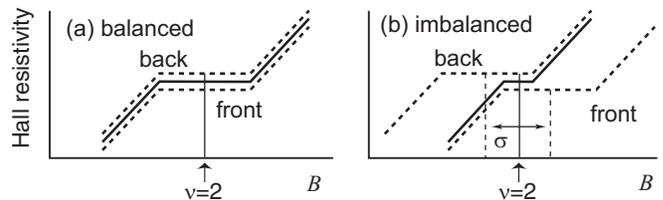}
\caption{Illustration of the F phase at $\protect\nu =2$. 
It consists of two monolayer $\protect\nu =1$ QH states in the front and back
layers. 
(a) At the balanced point, the $\nu =1$ Hall plateau (dotted lines) develops due to impurity effects in each layer, and so does the $\nu =2$ Hall plateau (thick line). (b) At the imbalanced point ($\sigma >0$), the number of electrons is more (less) than the number of flux quanta in the front (back) layer. In the front (back) layer with more (less) electrons, the Hall plateau (dotted line) would be generated with its center being located at a higher (lower) field, as indicated by a dotted vertical line. Then the $\nu =2$ Hall plateau (thick line) is generated in the region where the two $\nu =1$ Hall plateaux (dotted lines) in the front and back layers overlap.
}
\label{plateau}
\end{figure}

It is found that the experimentally determined F region has a finite width in $\sigma$. 
According to a theoretical work\cite{Ezawa_can}, however, the F phase is only stable
just at the balanced point for the ideal case without impurities. 
The origin of this discrepancy can be attributed to an impurity effect. 
In the F phase, the $\nu=1$ QH states are formed in both front and back layers.  Once impurities broaden the plateau width for each layer, 
the F phase appears in the overlap region of two $\nu=1$ QH states 
even if the density imbalance is made between the two layers.
This is illustrated in Fig.\thinspace\ref{plateau}.

We continue to analyze the phase diagram  in the $\sigma $-$n_{\mathrm{T}}$ plane [Fig.\thinspace\ref{sigma_nt}(b)]. We now focus on the minimum of $\Delta$ in Fig.\thinspace\ref{delta_sigma}.
We extract a set of values ($\sigma,n_{\mathrm{T}}$) that gives the minimum of $\Delta$, which are drawn as open diamonds in the phase diagram.
It is interesting that the boundary (open diamonds) does not coincide either with the CAF-F boundary or the CAF-S boundary.
It means that there exist two regions shown as II and IV in the CAF phase.
The II-IV boundary seems to be smoothly extrapolated to the I-II boundary point at $\sigma=0$ (dashed line) 
because no minimum was found at $n_{\mathrm{T}}=1.1\times 10^{11}\mathrm{cm}^{-2}$ 
in Fig.\thinspace\ref{delta_sigma}(a).

\begin{table}[t]
\begin{center}
\begin{tabular}{c|cccc}
\hline\hline
Region & \hspace{18pt} I \hspace{18pt} & \hspace{18pt} IV \hspace{18pt} & 
\hspace{18pt} II \hspace{18pt} & \hspace{18pt} III \hspace{18pt} \\ \hline
Phase & S & SCAF & FCAF  & F \\ \hline
$\frac{\partial\Delta}{\partial n_{\mathrm{T}}}$ & + & - & - & + \\ 
$\frac{\partial\Delta}{\partial\sigma}$ & + & + & - & - \\ \hline\hline
\end{tabular}%
\end{center}
\par
\caption{Summary of the notation of each phase and its properties against
the total density $n_{\mathrm{T}}$ and the density imbalance $\protect%
\sigma$.}
\end{table}

The identification of the regions II and IV is intriguing.
The characteristic features of the activation energy $\Delta $ are summarized as follows: As the density imbalance increases, $\Delta $ decreases in the
region II just like in the F phase, while $\Delta $ increases in the region IV
just as in the S phase. Recall that the intralayer (interlayer) interaction
destabilarizes (stabilarizes) the QH system against the density imbalance.
Besides the CAF phase emerges by the interplay between the spin and
pseudospin interactions, that is to say, between intralayer and interlayer interactions. Since it is natural that the behavior of the
activation energy is mainly controlled by the dominant interaction, the intralayer(interlayer) interaction is dominant in the region II (IV). 
Hence we call the regions II and IV the F-like CAF (FCAF) and the S-like CAF
(SCAF), respectively. The experimentally found II-IV region boundary in
Fig.\thinspace\ref{sigma_nt}(b) must represent the balanced point between the interlayer
and intralayer correlations in the two dimensional electron systems.


The structure of the spin and pseudospin has been calculated in each phase
\cite{Ezawa_can}, which is illustrated in Fig.\thinspace\ref{sigma_nt}(b). The magnitude
of the total spin $|S|$ (the pseudospin $|P|$) is maximal in the F phase (S phase), 
decreases in the CAF phase and finally vanishes in the S phase (F phase). 
It is well known\cite{Zheng,Das} that the spins are canted and make
an antiferromagnetic correlation between the two layers in the CAF phase.
The II-IV region boundary would be given when the ratio $|S|/|P|$ takes a
certain critical value depending on the tunneling and Zeeman gaps, though
its theoretical understanding is yet to be explored. We summarize various
properties of each region in Table.\thinspace 1.

\section{PHASE DIAGRAM IN THE $n_{\mathrm{T}}$-$\Delta_{\mathrm{SAS}}$ PLANE}

\begin{figure}[t]
\includegraphics[width=0.96\linewidth]{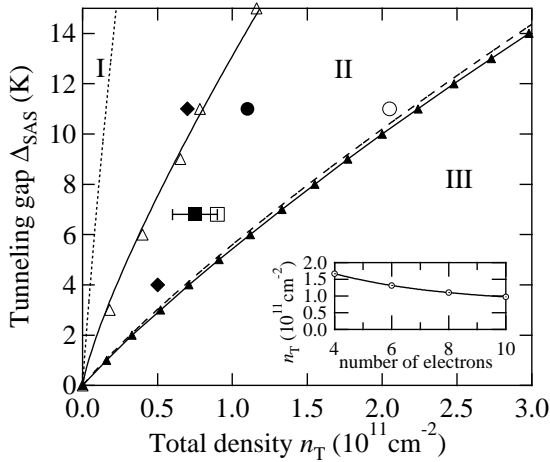}
\caption{Phase diagram in the $n_{\mathrm{T}}$-$\Delta_{\mathrm{SAS}}$ plane
at the balanced point. The solid and open circles are the S(I)-CAF(II) and CAF(II)-F(III) 
phase transition points derived from the data in Fig.\thinspace\ref{delta_nt}, respectively.
The open square is a re-interpreted  S-CAF phase boundary, 
while the error bar on the solid square shows that the CAF-F phase boundary must
be within this region,  as estimated from Ref.\,\onlinecite{Sawada_PRL}. 
The solid diamonds are re-interpreted S-CAF phase boundary determined from the SCAF-FCAF region boundary appeared in Ref.\,\onlinecite{Kumada}. See details in the text.
The dotted line and dashed line are the calculated S-CAF and CAF-F phase boundaries 
by the Hartree-Fock approximation. Open triangles and solid triangles
are the calculated S-CAF and CAF-F phase boundary points 
by an exact diagonalization method. 
Inset: Convergency of the ED method. The total density $n_{\mathrm{T}}$ 
at the CAF-F transition point vs. the number of
electrons to be diagonalized for $\Delta_{\mathrm{SAS}} = 11\,\mathrm{K}$.}
\label{nt_deltaSAS}
\end{figure}

Finally we construct the phase diagram in the $n_{\mathrm{T}}$-$\Delta _{%
\mathrm{SAS}}$ plane.  
The theoretical S-CAF phase boundary is calculated by the HF approximation from the equation: 
\cite{MacDonald_Rajaraman,Das,Ezawa_can}
$$\Delta_{\rm Z}=\sqrt{\Delta_{\rm SAS}(\Delta_{\rm SAS}-E_{\rm C})} ,$$
where $E_{\rm C}$ is the capacitance energy between two layers.
In the same way, the CAF-F phase boundary is determined from the equation: 
$$\Delta_{\rm Z}=\sqrt{\Delta_{\rm SAS}^2+\left(\frac{E_{\rm C}}{2}\right)^2}-\frac{E_{\rm C}}{2}.$$
They are shown in Fig.\thinspace\ref{nt_deltaSAS} by the dotted  and dashed lines, respectively. 
However, an ED study has suggested\cite{Schliemann} that the S-CAF
phase boundary in the HF approximation is not reliable. We carried out
an ED analysis of the phase diagram in the $n_{\mathrm{T}}$-$\Delta _{\mathrm{SAS%
}}$ plane, whose results are plotted as open triangles with solid lines in the
same figure. The convergency for the CAF-F phase boundary is quite good for
more than 6 particles to be diagonalized. On the other hand, calculated
points on the S-CAF boundary converge only gradually against 
the number of particles $N$ (see the inset
in Fig.\thinspace\ref{nt_deltaSAS}). We  display the asymptotic values for $%
N\rightarrow \infty $ in Fig.\thinspace\ref{nt_deltaSAS}. The agreement of the CAF-F
boundary between the HF approximation and the ED analysis is extremely good.
This is because the HF result is exact for the CAF-F boundary\cite{Ezawa_can}.

We mark the $n_{\mathrm{T}}$-$\Delta _{\mathrm{SAS}}$ phase diagram 
(Fig.\thinspace\ref{nt_deltaSAS}) with open and solid circles representing the present experimental results. The agreement of our experimental data points with the ED analysis is reasonably well but not perfect. The disagreement would be due to the finite width of quantum wells, which is neglected in theoretical calculations but is known to modify the Coulomb energy considerably. 

Here we review the experimental data given in Ref.\thinspace \onlinecite{Sawada_PRL}, 
where a sample with $\Delta _{\mathrm{SAS}}=6.7 \mathrm{K}$ was used. 
In view of our present understanding, the phase transition point at 
$n_{\mathrm{T}}=0.9\times 10^{11}\mathrm{cm}^{-2}$
in Fig.\thinspace 4 of this reference is interpreted as the CAF-F phase boundary. 
On the other hand, there is no definite data for the  S-CAF phase boundary. 
Nevertheless, Fig.\thinspace 3 of this reference tells us that
the QH state at $n_{\mathrm{T}}=0.6\times 10^{11}\mathrm{cm}^{-2}$ is in the S phase. 
Hence, the S-CAF phase boundary must be between $0.6\times 10^{11}\mathrm{cm}^{-2}$ 
and $0.9\times 10^{11}\mathrm{cm}^{-2}$.
We plot these points as squares in Fig.\thinspace\ref{nt_deltaSAS}.
We also re-analyzed the data based on Hall-plateau width measurements in Ref.\,\onlinecite{Kumada}. Though they did not find any phase transition points at $\sigma = 0$ for  two samples with $\Delta _{\mathrm{SAS}}=1 \mathrm{K}$ and $\Delta _{\mathrm{SAS}}=23 \sim 32 \mathrm{K}$, they did for two other samples with $\Delta _{\mathrm{SAS}}=4 \mathrm{K}$ and $\Delta _{\mathrm{SAS}}=11 \mathrm{K}$. They identified the  point that gives the minimum of the Hall plateau width for $\sigma$ at the fixed $n_\mathrm{T}$ as the F-S phase transition point. 
However, from our current understanding, the point which gives
the minimum of $\Delta$ against $\sigma$ should be interpreted as the SCAF-FCAF phase boundary. This point is re-interpreted as the S-CAF (S-FCAF) phase transition point due to the fact that the SCAF region seems to vanish at $\sigma=0$ in the present work. We also plot these two points as solid diamonds in Fig.\thinspace\ref{nt_deltaSAS}. The S-CAF phase boundary of the current data is slightly different from  the one determined from Ref.\,\onlinecite{Kumada}. One reason of this discrepancy of the S-CAF phase boundary is the method to determine the phase boundary. To analyze the S-CAF phase transition point in Ref.\,\onlinecite{Kumada}, we need the assumption that the SCAF phase vanishes at $\sigma=0$, and thus the FCAF-SCAF phase boundary coincides to the S-FCAF phase boundary. 
Another possibility  is different measuring methods. 
Though the Hall plateau width measurements and the activation energy measurements present qualitatively similar results, there is no reason that these two results are identical.
The activation energy measurement is more reliable.

\section{COMMENTS}

Carrying out  elaborate magnetotransport measurements,
we have established the existence of three phases in the $\nu =2$ bilayer QH system. 
Our new finding is an intrinsic structure of the CAF phase represented 
by the FCAF and SCAF regions in the imbalanced density configuration.


We make comments on related theoretical works. Brey {\it et al.} predicted that there arise new coherent phases  for large  layer density imbalance\cite{Brey}. For large bias voltage the CAF phase would become the coherent canted (CC) phase. 
However this CC phase is only stable at zero tunneling limit $\Delta_{\rm SAS}=0$. Since our sample has a large tunneling energy, it is difficult to identify the region IV as the CC phase. 
Demler {\it et al.} suggested that there exist the Bose-glass phase due to nonzero disorder effect\cite{Demler_Bose}. Though their theoretical calculation was performed only in the balanced density configuration, it would be worthwhile to try to identify the region II (IV) with a Bose-glass phase made of domains of the CAF phase surrounded by domains of the F (S) phase. However, a theoretical study of the Bose-glass phase in the imbalanced configuration will be indispensable for further discussions.

\begin{acknowledgments}
We are grateful to T. Saku for growing the heterostructures, and K. Muraki
for fruitful discussions. We also thank to T. Nakajima for many fruitful
advises for theoretical calculations. This research was supported in part by
Grants-in-Aid for the Scientific Research and Technology of Japan (Nos.
14010839,14340088) and a 21st Century COE Program Grant of the International
COE of Exploring New Science Bridging Particle-Matter Hierarchy from the
Ministry of Education, Culture, Sports, Science. 
A part of this work was performed at the clean room facility of the Center for Interdisciplinary Research of Tohoku University. 
\end{acknowledgments}

\bibliography{can2}

\end{document}